\documentclass[11pt]{article}
\newcommand{\Comment}[1]{{}}
\usepackage{amsfonts,amsthm,amsmath,amssymb,slashed}
\usepackage[textwidth = 430 pt, textheight = 630 pt]{geometry}

\usepackage{color}
\definecolor{MyDarkBlue}{rgb}{0.15,0.15,0.45}
\usepackage[linktocpage=true]{hyperref}
\hypersetup{
colorlinks=true,
citecolor=MyDarkBlue,
linkcolor=MyDarkBlue,
urlcolor=MyDarkBlue,
pdfauthor={Horatiu Nastase and Constantinos Papageorgakis},
pdftitle={Dimensional reduction of the ABJM model},
pdfsubject={hep-th}
}

\usepackage[numbers,sort&compress]{natbib}
\usepackage{hypernat}

\newcommand\ignore[1]{}
\def\one{{\,\hbox{1\kern-.8mm l}}}

\def\Tr{{\rm Tr\, }}

\newcommand{\SO}{\mathrm{SO}} 
\newcommand{\SU}{\mathrm{SU}} \newcommand{\U}{\mathrm{U}}
\newcommand{\ie}{\emph{i.e.}\:}
 \newcommand{\pd}{\partial}

\def\a{\alpha}\def\b{\beta}

\def\d{\partial}

\newcommand{\cp}{\Cset \mathrm{P}}
\newcommand{\Cset}{{\,\,{{{^{_{\pmb{\mid}}}}\kern-.45em{\mathrm C}}}}}

\newcommand{\be}{\begin{equation}}
\newcommand{\bea}{\begin{eqnarray}}

\newcommand{\ee}{\end{equation}}
\newcommand{\eea}{\end{eqnarray}}

\newcommand{\nn}{\nonumber}

\begin{document}

\renewcommand{\thefootnote}{\fnsymbol{footnote}}

\makeatletter
\@addtoreset{equation}{section}
\makeatother
\renewcommand{\theequation}{\thesection.\arabic{equation}}

\rightline{}
\rightline{}
   \vspace{1.8truecm}

\vspace{15pt}

%%%%%%%%%%%%%%%%%

\centerline{\LARGE \bf{\sc Dimensional reduction of the ABJM model}} 
 \vspace{1truecm}
\thispagestyle{empty} \centerline{
    {\large \bf {\sc Horatiu Nastase${}^{a,}$}}\footnote{E-mail address: \Comment{\href{mailto:nastase@ift.unesp.br}}{\tt 
    nastase@ift.unesp.br}} {\bf{\sc and}}
    {\large \bf {\sc Constantinos Papageorgakis${}^{b,}$}}\footnote{E-mail address:
                                \Comment{\href{mailto:costis.papageorgakis@kcl.ac.uk}}{\tt costis.papageorgakis@kcl.ac.uk}}
                                                           }

\vspace{1cm}
\centerline{{\it ${}^a$ 
Instituto de F\'{i}sica Te\'{o}rica, UNESP-Universidade Estadual Paulista}} \centerline{{\it 
R. Dr. Bento T. Ferraz 271, Bl. II, Sao Paulo 01140-070, SP, Brazil}}

\vspace{.8cm}
\centerline{{\it ${}^b$ 
Department of Mathematics, King's College London}} \centerline{{\it The Strand, London WC2R 2LS, UK}}

\vspace{2truecm}

%%%%%%%%%%%%%%%%%
\thispagestyle{empty}

\centerline{\sc Abstract}

\vspace{.4truecm}

\begin{center}
  \begin{minipage}[c]{380pt}{\noindent We dimensionally reduce the ABJM model, obtaining a       two-dimensional theory that can be thought of as a `master action'. This encodes information       about both T- and S-duality, \ie describes fundamental (F1) and D-strings (D1) in 9 and 10       dimensions.  The Higgsed theory at large VEV, $\tilde v$, and large $k$ yields D1-brane       actions in 9d and 10d, depending on which auxiliary fields are integrated out. For $N=1$       there is a map to a Green-Schwarz string wrapping a nontrivial circle in $\mathbb       C^4/\mathbb Z_k$.  }
\end{minipage}
\end{center}

\vspace{.5cm}

\setcounter{page}{0}
\setcounter{tocdepth}{2}

\newpage

%\tableofcontents
\renewcommand{\thefootnote}{\arabic{footnote}}
\setcounter{footnote}{0}

\linespread{1.1}
\parskip 4pt

{}~
{}~

\section{Introduction}

The Aharony-Bergman-Jafferis-Maldacena (ABJM) model \cite{Aharony:2008ug} has received a lot of attention lately, as it captures the dynamics of multiple M2-branes in a particular M-theory background.  Whereas the Bergshoeff-Sezgin-Townsend (BST) action \cite{Bergshoeff:1987cm} for a single membrane has no gauge fields, is generically nonconformal and contains the membrane tension parameter $T_2$, the ABJM theory is a $\U(N)\times\U(N)$, conformal, Chern-Simons-matter gauge theory at level $k$ with $\mathcal{N}=6$ supersymmetry, corresponding to the IR limit of $N$ M2-branes on a $\mathbb C^4/\mathbb Z_k$ singularity. The interest in the ABJM model arises both from being the first  example of a multiple membrane theory, as well as from the fact that it provides a new direction for $\textrm{AdS}_4/\textrm{CFT}_3$, being dual to string theory in the $\textrm{AdS}_4\times \cp^3$ background at large $k$.\footnote{The near-horizon geometry for M2-branes on $\mathbb C^4/\mathbb Z_k$ is   $\textrm{AdS}_4\times S^7/\mathbb Z_k$. The orbifold action is such that $S^1/\mathbb   Z_k\hookrightarrow S^7/\mathbb Z_k\stackrel{\pi}{\rightarrow} \cp^3$, so the geometry reduces to $\textrm{AdS}_4\times \cp^3$  in the large-$k$ limit.}

It is well known that, after double dimensional reduction, the BST action yields the Green-Schwarz (GS) description of a fundamental string (F1) \cite{Duff:1987bx}. One could also consider reducing on two, instead of one, circles: The M2 can first be compactified on a worldvolume direction down to an F1-string in 10d and then on a transverse direction to a type IIA F1 in 9d. But if instead one first reduces on a transverse and then on a parallel direction, the result is a type IIA D1 in 9d. The two procedures are related by an S-duality transformation, after also having implemented a T-duality to IIB configurations in 10d. 

In the same spirit it is natural to expect that the dimensionally-reduced ABJM model should  also be related to a multiple fundamental string action. An immediate problem with that assumption is that there exists no such known example. Moreover one would not expect it to contain gauge fields, obtained from the ABJM Chern-Simons gauge fields, which would be more in line with having a theory of D1-branes. Finally, there is also an additional effective compactification occurring for large $k$ in the spirit of \cite{Mukhi:2008ux,Distler:2008mk}, which however seems to commute with the na\"ive dimensional reduction and hence contradicts the intuition of the two-circle compactification described above.

In this note we will analyse the dimensional reduction of the ABJM model in more detail and argue that the resulting theory can be interpreted as a `master action' that encodes information for both T- and S-duality: Depending on the variables used and the energy regime one is interested in, we obtain either multiple F- or D-strings in 9d or 10d. There exists some related work \cite{Santos:2008ue,Franche:2008hr}, particularly pertaining to the dimensional reduction of the Bagger-Lambert-Gustavsson (BLG) model \cite{Bagger:2007jr,Gustavsson:2007vu}, although in our opinion the interpretation of the final action has not been explored in the same fashion. \Comment{The BLG model is an $\mathcal N=8$ CS-matter gauge theory based on real 3-algebras. Its only known consistent realisation, in terms of the ${\cal A}_4$ 3-algebra, is equivalent to the $\SU(2)\times\SU(2)$  ABJM model \cite{VanRaamsdonk:2008ft}. There is currently no known String/M-theory interpretation for this theory.\footnote{However for   connections between $\SU(N)\times\SU(N)$ and $\U(N)\times\U(N)$ ABJM theories see   \cite{Lambert:2010ji}.}}

The rest of this paper is organised as follows. We perform the dimensional reduction, showing in Section 2.1 that in the presence of a large VEV the Higgsed theory reduces to either the D1-brane action in 9d or the D1-brane action in 10d, with the two related by T-duality. We also calculate the action for $N=1$ in terms of a particular set of variables. In Section 3 we proceed to exhibit a transformation that turns the $N=1$ ABJM model into the usual BST action. The same transformation turns our reduced action for $N=1$ into a GS-string action. We finally interpret the results, showing in particular that the latter correctly corresponds to a string in a $\mathbb C^4/\mathbb Z_k$ background for general parameters.

\section{Dimensional reduction}

The ABJM action \cite{Aharony:2008ug}, corresponding to the IR limit of $N$ M2-branes at an $\mathbb R^{2,1}\times \mathbb C^4/\mathbb Z_k$ singularity, is given by
\bea
S_{ABJM}&=&\int d^3 x\Big[\frac{k}{4\pi}\epsilon^{\mu\nu\lambda}{\rm Tr}\Big(A_\mu^{(1)}\d_\nu A_\lambda^{(1)}+\frac{2i}{3}
A_\mu^{(1)} A_\nu^{(1)} A_\lambda^{(1)}-A_\mu^{(2)}\d_\nu A_\lambda^{(2)}-\frac{2i}{3}
A_\mu^{(2)} A_\nu^{(2)} A_\lambda^{(2)}\Big)\cr
&&-{\rm Tr}\Big( D_\mu C^\dagger_ID^\mu C^I\Big) -i {\rm Tr}\Big(\psi^{I\dagger}\gamma^\mu D_\mu \psi_I\Big) \nonumber\\
&&\left.+\frac{4\pi^2}{3k^2}{\rm Tr}\left(C^IC_I^\dagger C^JC_J^\dagger
C^KC_K^\dagger+C^\dagger_IC^IC^\dagger_JC^JC^\dagger_KC^K\right.\right.\nonumber\\ 
&&\left.\left.+4 C^IC^\dagger_JC^KC^\dagger_IC^JC^\dagger_K-6 C^IC^\dagger_JC^JC^\dagger_IC^KC^\dagger_K\right)\right.\nonumber\\
&&\left.+\frac{2\pi i}{k}{\rm Tr}\left(C^\dagger_IC^I\psi^{J\dagger}\psi_J-\psi^{\dagger J}C^IC^\dagger_I
\psi_J-2 C^\dagger_IC^J\psi^{\dagger I}\psi_J+2\psi^{\dagger J}C^IC^\dagger_J\psi_I\right.\right.\nonumber\\
&&\left.\left. +\epsilon^{IJKL}\psi_IC^\dagger_J\psi_KC^\dagger _L-\epsilon_{IJKL}\psi^{\dagger I}C^J
\psi^{\dagger K}C^L\right)\right]\;,
\label{abjmaction}
\eea
where the Lorentz index $\mu = 0,1,2$ and the R-symmetry index $I=1,...,4$ in $\SU(4)$.

There also exists a maximally supersymmetric massive deformation \cite{Hosomichi:2008jb, Gomis:2008vc} where one splits the scalars as $C^I=(R^\a,Q^{\alpha})$, with $\alpha = 1,2$. Then the mass deformation changes the potential to 
\be
V=|M^{ \alpha}|^2+|N^\a|^2\label{potential}\;,
\ee
where 
\bea
M^{ \alpha}&=& \mu Q^{ \alpha} +\frac{2\pi}{k}(2Q^{[ \alpha }Q^\dagger_{\b} Q^{\b ]}+R^{\b} R^\dagger_{\b} Q^{\a}-Q^{\a} R^\dagger_\b R^\b
+2Q^{\b} R^\dagger_\b R^\a)\nonumber\\
N^\a&=&-\mu R^\a +\frac{2\pi}{k}(2R^{[\a}R^\dagger_\b R^{\b ]}+Q^\b Q^\dagger_{\b} R^\a-R^\a Q^\dagger_{\b} Q^{\b}
+2R^\b Q^\dagger_{\b} Q^{\a})\label{potterms}\;.
\eea
In addition, the potential involves a mass term $\mu$ for the fermions.

In order to dimensionally reduce the above on a circle of radius $R$, we choose the standard ansatz by dropping the dependence of all fields on the circle direction $y$, where we split $x^\mu=x^i, y$, with $i=0,1$.\footnote{Since we are compactifying on a circle by keeping all the fields and just   dropping the circle dependence, we have automatically obtained a consistent truncation: The spherical harmonics are trivial (Fourier modes), so there are no terms linear in the massive   (dropped) fields in the action (a massive Fourier mode $e^{2\pi iny/R}$ needs at least another   massive mode $e^{-2\pi i n/R}$ to give a nonzero result after integration).} We further need to rescale the fields by powers of the radius $R$ to obtain the canonical dimension in 2d. This leads to the following ansatz
\be
C^I=\frac{1}{\sqrt{R}}\tilde{C}^I(\vec x);\;\;\;
\psi=\frac{1}{\sqrt{R}}\tilde{\psi}(\vec x);\;\;\;
A_i^{(1,2)}=A_i^{(1,2)}(\vec x);\;\;\;
A_y^{(1,2)}=\frac{1}{R}\phi_{1,2}(\vec x)\;.
\ee
The covariant derivatives are 
\bea
&&D_iC^I=\d_i C^I-iA_i^{(1)}C^I+iC^IA_i^{(2)}\cr
&&D_y C^I=i\frac{1}{R}(C^I\phi_2-\phi_1 C^I)
\eea
and similarly for  the fermions. From the `pure' (undeformed) ABJM action (\ref{abjmaction}) we then get the dimensionally reduced action
\bea
S_{2d}&=&\int d^2x \Big[\frac{k}{4\pi}\epsilon^{ij}\Tr\Big(\phi_1F^{(1)}_{ij}-\phi_2F^{(2)}_{ij}\Big) 
-{\rm Tr}\Big( D_i \tilde C^\dagger_ID^i \tilde C^I\Big) -i {\rm Tr}\Big(\tilde \psi^{I\dagger}\gamma^i D_i 
\tilde \psi_I\Big) \nn\\
&&+\frac{4\pi^2}{3k^2R^2}{\rm Tr}\Big(\tilde C^I\tilde C_I^\dagger \tilde C^J\tilde C_J^\dagger
\tilde C^K\tilde C_K^\dagger+\tilde C^\dagger_I\tilde C^I\tilde C^\dagger_J\tilde C^J\tilde C^\dagger_K\tilde C^K\nonumber\\ 
&&+4 \tilde C^I\tilde C^\dagger_J\tilde C^K\tilde C^\dagger_I\tilde C^J\tilde C^\dagger_K
-6 \tilde C^I\tilde C^\dagger_J\tilde C^J\tilde C^\dagger_I\tilde C^K\tilde C^\dagger_K\Big)\nonumber\\
&&+\frac{2\pi i}{kR}{\rm Tr}\Big(\tilde C^\dagger_I\tilde C^I\tilde \psi^{J\dagger}\tilde \psi_J-\tilde \psi^{\dagger J}\tilde C^I
\tilde C^\dagger_I\tilde \psi_J
-2 \tilde C^\dagger_I\tilde C^J\tilde \psi^{\dagger I}\tilde\psi_J+2\tilde \psi^{\dagger J}\tilde C^I\tilde C^\dagger_J
\tilde \psi_I\nonumber\\
&& +\epsilon^{IJKL}\tilde \psi_I\tilde C^\dagger_J\tilde \psi_K\tilde C^\dagger _L-\epsilon_{IJKL}\tilde \psi^{\dagger I}\tilde C^J
\tilde \psi^{\dagger K}\tilde C^L\Big)\cr
&& +R^{-2}\Tr\Big((\tilde C^I\phi_2-\phi_1\tilde C^I)(\tilde C^\dagger_I\phi_1-\phi_2\tilde C^\dagger_I)\Big)\cr
&& +R^{-1}\Tr\Big(\tilde \psi^{I\dagger}\gamma_3(\tilde \psi_I\phi_2-\phi_1\tilde\psi_I)\Big)\Big]\label{redaction}\;,
\eea
where
\be
F^{(1,2)}_{ij}=\d_iA_j^{(1,2)}-\d_jA^{(1,2)}_i+iA_i^{(1,2)}A_j^{(1,2)}-iA_j^{(1,2)}A_i^{(1,2)}\;.
\ee
The fields $\phi_{1,2}$ are auxiliary (nonpropagating) and as a result could  be eliminated from the action. 

In order to get a feeling for the general case, we first set $\phi_2=0$. This is not a solution, \ie a consistent truncation, so the following is just for purposes of illustration. The bosonic 
$\phi_1$ action reduces to 
\be\label{phi1action}
\int d^2 x\Big[\frac{\sqrt{2}}{2 g R}\Tr(\epsilon^{ij}\phi_1 F_{ij}^{(1)})-R^{-2}\Tr((\phi_1)^2\tilde C^I\tilde C^\dagger_I)\Big]\;,
\ee
where we have defined
\be
g=\frac{2\pi\sqrt{2} }{kR}\;.
\ee
Solving for $\phi_1$
\be
\phi_1=\frac{R\sqrt{2}}{4 g}\epsilon^{ij}F_{ij}^{(1)}(\tilde C^I\tilde C^\dagger_I)^{-1}
\ee
and replacing in (\ref{phi1action}), while using that in 2d $(\epsilon^{ij}F_{ij})^2=-2F_{ij}F^{ij}$, we obtain
the kinetic term
\be
-\frac{1}{4 g^2}\int d^2x \Tr \Big[F_{ij}^{(1)}F^{(1)ij}(\tilde C^I\tilde C^\dagger_I)^{-1}\Big]\;.
\ee
This is the standard kinetic term for the gauge field $A_i^{(1)}$, with a nonpolynomial scalar field dressing-factor.

However, when we include $\phi_2$, we get the equations
\bea
&& \frac{R\sqrt{2} \epsilon^{ij}F_{ij}^{(1)}}{2 g}=2\phi_1(\tilde C^I\tilde C^\dagger_I)-2C^I\phi_2\tilde C^\dagger_I
-R\tilde \psi_{I\a}(\tilde\psi^{I\dagger}\gamma_3)^\a\nn\\
&&-\frac{R \sqrt{2}\epsilon^{ij}F_{ij}^{(2)}}{2 g}=2\phi_2(\tilde C^\dagger_I\tilde C^I)-2C^\dagger_I\phi_1\tilde C^I
-R(\tilde\psi^{I\dagger}\gamma_3)^\a\tilde\psi_{I\a}\;,
\eea
where we have explicitly written out the spinor indices $\alpha=1,2$. From the above one can derive an equation for $\phi_1$
\bea
\phi_1(\tilde C^I\tilde C^\dagger_I)-\tilde C^K\tilde C^\dagger_J\phi_1\tilde C^J(\tilde C^\dagger_I \tilde C^I)^{-1}\tilde C^\dagger_K &=& \frac{R\sqrt{2}}{4 g}\epsilon_{ij}\Big[F_{ij}^{(1)}-\tilde C^KF_{ij}^{(2)}(\tilde C^\dagger_I\tilde C^I)^{-1}\tilde C^\dagger_K\Big]\cr
&&-\frac{R}{2} \tilde C^K(\tilde \psi^{J\dagger}\gamma_3)^\a\tilde\psi_{J\a}(\tilde C^\dagger_I\tilde C^I)^{-1}\tilde C^\dagger_K\cr
&&+\frac{R}{2}\tilde \psi_{I\a}(\tilde \psi^{I\dagger}\gamma_3)^\a
\eea
that we cannot solve further. In principle, the solutions for $\phi_1,\phi_2$ should be then substituted back into the action
\bea
S_{2d}&=&\int d^2x \Big[\frac{\sqrt{2}}{2 g R}\Tr\Big(\phi_1\epsilon^{ij}F_{ij}^{(1)}-\phi_2\epsilon^{ij}F_{ij}^{(2)}\Big)\cr
&&\qquad-R^{-2}
\Tr\Big((\phi_1)^2\tilde C^I\tilde
C^\dagger_I+(\phi_2)^2\tilde C^\dagger_I\tilde C^I-2\phi_1\tilde C^I\phi_2\tilde C^\dagger_I\Big)\cr
&&\qquad +R^{-1}\Tr\Big(\phi_2\tilde \psi^{I\dagger}\gamma_3\tilde \psi_I+\phi_1\tilde \psi_{I\a}(\tilde \psi^{I\dagger}\gamma_3)^\a\Big)\Big]
\eea
and added to the $\phi_i$-independent part.

Note that the fields are massless, but the action is nonconformal since $g$ has dimensions of mass, as expected for the Yang-Mills coupling in 2d.  It is also important to mention that the gauge fields that have been obtained are still nonpropagating, as the YM kinetic term in 2d has $d-2=0$ degrees of freedom, and thus there is no contradiction with the counting of degrees of freedom before and after the reduction (the scalars remain scalars and the fermions do not lose degrees of freedom when going from 3d to 2d).

Dimensionally reducing the mass-deformed ABJM theory is trivial: The mass deformation only  affects the scalar potential and gives mass to the fermions, so these terms remain unaffected by going down to 2d.  Similarly, the gauge field kinetic terms are the same as those for the undeformed ABJM theory, except for the fact that the supersymmetric vacuum is now the fuzzy sphere as in \cite{Gomis:2008vc,Nastase:2009ny,Nastase:2009zu,Nastase:2010uy}.

\subsection{Higgsing the reduced theory}\label{higgssection}

We next investigate the vacuum structure of the 2d theory. The VEV $\langle\tilde C^I\rangle=\tilde v \delta^I_1  \one_{N\times N}$, with the rest of the fields set to zero, is a solution of the equations of motion. Expanding the theory (\ref{redaction}) around this vacuum, and fixing for the 
moment the scalars to their VEV values, we obtain 
\be\label{vevaction}
S=\int d^2 x\Big[-i \Tr(\tilde \psi^{I\dagger}\gamma^iD_i\tilde \psi_I)-\tilde v^2\Tr(A_i^{(1)}-A_i^{(2)})^2  \Big]+S_{\phi}\;,
\ee
where
\begin{multline}
S_{\phi}=\int d^2x
\Big[\frac{\sqrt{2}}{2 gR}\Tr(\phi_1\epsilon^{ij}F_{ij}^{(1)}-\phi_2\epsilon^{ij}F_{ij}^{(2)})+R^{-1}\Tr(\phi_2\tilde \psi^{I
\dagger}\gamma_3\tilde\psi_I+\phi_1\tilde\psi_{I\a}(\tilde \psi^{I\dagger}\gamma_3)^\a)\\-\tilde v^2R^{-2}\Tr(\phi_1-\phi_2)^2\Big]\;.
\end{multline}
Varying with respect to $\phi_1,\phi_2$ we then get the following constraints:
\bea
\phi_1-\phi_2&=&\frac{R\sqrt{2}}{4\tilde v^2 g}\epsilon^{ij}F^{(1)}_{ij}+\frac{R}{2\tilde v^2}
\tilde \psi_{I\a}(\tilde \psi^{I\dagger}\gamma_3)^\a\cr
\epsilon^{ij}F_{ij}^{(2)}&=&\epsilon^{ij}F_{ij}^{(1)}+\sqrt{2} g(\tilde \psi_{I\a}(\tilde \psi^{I\dagger}\gamma_3)^\a+\tilde 
\psi^{I\dagger}\gamma_3 \psi_{I})\;.
\eea
This implies that  $F^{(2)}_{ij}$ is  determined  in terms of $F^{(1)}_{ij}$, as is $\phi_1-\phi_2$, although $\phi_1+\phi_2$ is still free. Nevertheless, substituting back in the auxiliary field action we get that $\phi_1+\phi_2$ also disappears from the action to give
\be\label{YMaction}
S =\int d^2 x\Big[-i \Tr(\tilde \psi^{I\dagger}\gamma^iD_i\tilde \psi_I)-\tilde v^2\Tr(A_i^{(1)}-A_i^{(2)})^2 \Big]+S_{\phi}\;,
\ee
with
\be
S_{\phi}=\int d^3 x \Big[-\frac{1}{4 g^2\tilde v^2}\Tr(F_{ij}^{(1)}F^{(1)\;ij})
+\frac{\sqrt{2}}{4 g\tilde v^2}\Tr[\epsilon^{ij}F_{ij}^{(1)}\tilde \psi_{I\a}
(\tilde \psi^{I\dagger}\gamma_3)^\a]+\frac{1}{4\tilde v^2}\Tr[\tilde \psi_{I\a}(\tilde \psi^{\dagger I}\gamma_3)^\a]^2\Big]\;.
\ee
The first term in the above looks like a 2d YM kinetic term, with coupling   
\be
g_{YM}\equiv g \tilde v =\frac{2\pi \sqrt{2}\tilde v}{kR}\;.\label{couplingv}
\ee

Note that half of the gauge fields were fixed by the VEV. Moreover, if one chose to keep the scalar field fluctuations the second gauge field would also have to appear in a `kinetic term', multiplied by said fluctuations. In that event it is best to think of the latter as an interaction term, with the constraints fixing both $\phi_1$ and $\phi_2$ but the second gauge field remaining unfixed.

In the mass-deformed case, one keeps a kinetic term involving both gauge fields. This fixes both $\phi_1$ and $\phi_2$ even when expanding the theory around the fuzzy sphere vacuum and keeping only leading terms by setting their fluctuations to zero.

\subsubsection{Higgsed action at large $\tilde v$ and D1-brane in 9d}

We just saw that by eliminating the $\phi_i$'s in the absence of scalar fields one gets a nontrivial action with a 2d YM kinetic term. This renders it in principle compatible with a D1-brane interpretation. In the following we would like to show that at large $\tilde v$ and large $k$, as imposed by the finiteness of (\ref{couplingv}) with  $\tfrac{\tilde v}{k}=\textrm{fixed}$, we will obtain a D1-brane low-energy theory in 9d flat space. 

We begin by defining
\bea
\tilde B_y=\phi_1-\phi_2\;, && \tilde Q_y=\phi_1+\phi_2\cr
B_i=\tfrac{1}{2}(A_i^{(1)}-A_i^{(2)})\;, && Q_i=\tfrac{1}{2}(A_i^{(1)}+A_i^{(2)})\cr
F^B_{ij}=\tfrac{1}{2}(F_{ij}^{(1)}-F_{ij}^{(2)})\;, && F_{ij}=\tfrac{1}{2}(F^{(1)}_{ij}+F^{(2)}_{ij})\;,
\eea
which implies
\bea
F_{ij}&=&\d_i Q_j-\d_j Q_i+i[Q_i,Q_j]+i[B_i,B_j]\cr
F_{ij}^B&=& \tilde D_i B_j-\tilde D_j B_i\cr
\tilde D_i&\equiv &\d_i +i[Q_i,\cdot\;]\label{fdefs}\;.
\eea
The $\phi_i$-dependent terms in the action (\ref{redaction}) can then be rewritten as 
\begin{multline}
S_\phi=\int d^2 x\Big[\frac{k}{4\pi}\epsilon^{ij}\Tr\Big(\tilde B_yF_{ij}+\tilde Q_yF^B_{ij}\Big)+\frac{1}{2R}\Tr\Big(\tilde \psi^{I\dagger}\gamma_3([\tilde \psi_I,\tilde Q_y]-\{\tilde \psi_I,\tilde B_y\})\Big)\\
+\frac{1}{4R^2}\Tr\Big([\tilde C^I,\tilde Q_y]-
\{\tilde C^I,\tilde B_y\}\Big)\Big([\tilde C^\dagger_I,\tilde Q_y]+\{\tilde C_I^\dagger,\tilde B_y\}\Big)\Big]\label{sphi}\;.
\end{multline}
We expand the scalars around the VEV solution as 
\bea
\tilde C^1=\tilde v +\rho +i\sigma\;, && \rho=\rho_0+i\rho_a T^a\cr
\tilde C^{I'}=X^{I'}+iX^{I'+4} \;, && \sigma=\sigma_0+i\sigma_aT^a\cr
X^{A''}=X^{A''}_0+iX^{A''}_aT^a\;, &&
\eea
where $T^a$ are $\SU(N)$ generators, $I'=2,3,4$, $A''=(I',I'+4)$ and the subscript $0$ indicates the trace part.

The $\phi_i$-dependent action (\ref{sphi})  becomes to leading order in $\tilde v$
\begin{multline}\label{leading}
S_{\phi}=\int d^2 x\Big[\frac{k}{4\pi}\epsilon^{ij}\Tr\Big(\tilde B_yF_{ij}+\tilde Q_yF^B_{ij}\Big)+\frac{1}{2R}\Tr\Big(\tilde \psi^{I\dagger}\gamma_3([\tilde \psi_I,\tilde Q_y]-\{\tilde \psi_I,\tilde B_y\})\Big)\\
-\frac{\tilde v^2}{R^2}\Tr\Big(\tilde 
B_y^2\Big)+{\cal O}(\tilde v)\Big]\;,
\end{multline}
\ie  remains independent of the scalars.

In exact analogy to the 3d case \cite{Mukhi:2008ux,Pang:2008hw},\footnote{The calculation is identical and we will hence omit it at this stage. See also \cite{Chu:2010fk}.} we obtain for the scalar potential
\be
\frac{4\pi^2}{3k^2R^2}V_6(\tilde C^I)= -\frac{ 4\pi^2\tilde  v^2}{k^2R^2}\Tr[ X^{A'}, X^{B'}]^2+{\cal O}(\tilde v)
=\frac{g_{YM}^2}{2}\Tr [X^{A'}, X^{B'}]^2(1+{\cal O}(\tfrac{1}{\tilde v}))
\ee
where $A'=2,...,8$ and $X^5\equiv \sigma$. For the fermionic potential we have
\begin{multline}
\frac{i g \tilde v }{\sqrt 2}V_{ferm}=-\frac{i g_{YM}}{\sqrt{2}}\Tr\Big[2\rho(\tilde \psi^{\dagger J}\tilde \psi_J-\tilde \psi_J\tilde \psi^{\dagger J})
-2\tilde \psi_1(\tilde C^\dagger_I\tilde \psi^{\dagger I}-\tilde \psi^{\dagger I}\tilde C_I^\dagger)
+2\tilde \psi^{\dagger 1}(\tilde C^I\tilde \psi_I-\tilde \psi_I\tilde C^I)\\
+2\epsilon^{I'J'K'}\tilde \psi_{I'}\tilde C^\dagger_{J'}\tilde \psi_{K'}
-2\epsilon_{I'J'K'}\tilde \psi^{\dagger I'}\tilde C^{J'}\tilde \psi^{\dagger K'}\Big](1+{\cal O}(\tfrac{1}{\tilde v}))
\end{multline}
and one similarly  obtains the anticipated SYM Yukawa term
\be
-\frac{1}{2}f^{abc}X_a^{A'}\bar\psi_b\Gamma^{A'}\psi_c+{\cal O}(\tfrac{1}{\tilde v})
\ee
by rearranging the fermions into $\SO(1,1)\times \SO(7)$ spinors, again in direct analogy with \cite{Pang:2008hw}. Note that the scalar $\rho_a$ (the real part of $\tilde C^1$) does not appear in either the final bosonic or  fermionic potentials, as was also the case in 3d. 

We now move on to eliminate the auxiliary fields $\tilde B_y$ and $\tilde Q_y$ from the action $S_\phi$. Varying (\ref{leading}) with respect to $\tilde B_y$ and $\tilde Q_y$, we obtain
\bea
\tilde B_y&=&\frac{\sqrt{2}R}{4g\tilde v^2}\epsilon^{ij}F_{ij}-\frac{R}{4\tilde v^2}(\tilde \psi_{I\a}(\tilde \psi^{I\dagger}\gamma_3)^\a+
\tilde \psi^{I\dagger}\gamma_3\tilde \psi_I)\cr
\epsilon^{ij}F_{ij}^B&=&\frac{2\pi}{kR}(\tilde \psi_{I\a}(\tilde \psi^{I\dagger}\gamma_3)^\a+
\tilde \psi^{I\dagger}\gamma_3\tilde \psi_I)\label{bconstraints}
\eea
and substituting back we find
\bea
S_{\phi}&=&\int d^2 x\Big[-\frac{1}{4g_{YM}^2}F_{ij}F^{ij}-\frac{\sqrt{2}}{8g\tilde v^2}\epsilon^{ij}F_{ij}(\tilde \psi^{I}_\alpha(\tilde 
\psi^{I\dagger}\gamma_3)^\a+\tilde \psi^{I\dagger}\gamma_3\tilde \psi_I)\cr
&&-\frac{1}{16\tilde v^2}(\tilde \psi^{I}_\alpha(\tilde \psi^{I\dagger}\gamma_3)^\a+
\tilde \psi^{I\dagger}\gamma_3\tilde \psi_I)^2\Big](1+{\cal O}(\tfrac{1}{\tilde v}))\cr
&\rightarrow& -\frac{1}{4g_{YM}^2}\int d^2 x F_{ij}F^{ij}+
{\cal O}(\tfrac{1}{\tilde v})\;,
\eea
that  is only the $F_{ij}$ YM kinetic term survives  in the large $\tilde v$ limit. We note that in the above there is no YM kinetic term for $F^B_{ij}$ and that to ${\cal O}(1)$ in $S_\phi$, the $\tilde Q_y$ field drops out from the action, despite $\tilde Q_y$ not being fixed to leading order.

Apart from the $\phi_i$-dependent part of the action, one also needs to take into consideration the $\tilde v^2 B_i B^i$ mass-terms coming from the covariant kinetic term for the scalars, $|D_i\tilde C^I|^2$. Even though $B_i$ is an adjoint field under $Q_i$, it is itself the gauge field for a shift symmetry that acts as
\bea
B_i &\to& B_i - \tilde D_i \lambda\cr
W &\to& W + \alpha \lambda\;,
\eea
with $W = \sigma_0 + i \rho_a T^a$ \cite{Chu:2010fk} and $\alpha$ an appropriate combination of the parameters of the theory. In order to proceed we observe that in the large-$k$ limit the $\tilde Q_y$ constraint becomes $\epsilon^{ij}F_{ij}^B=0$, which for topologically trivial fields is solved by a pure gauge condition
\be\label{puregauge}
B_i=\tilde D_i\lambda\;.
\ee
One might be tempted to think of this as a trivial solution but this is not the case. In fact, this is just a signal of the ordinary Higgs mechanism where the gauge field `eats' the Goldstone boson to become massive. Indeed, in 2d the YM gauge field is nondynamical, while a massive (Proca) vector field with Lagrangian density
\be
-\frac{1}{4g_{YM}^2}F_{ij}^2-m^2A_i^2
\ee
has one dynamical degree of freedom. For the case at hand there is no YM kinetic term but we still have a mass for $B_i$ from the scalar kinetic term, which renders it dynamical. In the final step, we substitute the single dynamical mode of $B_i$ through (\ref{puregauge}), in effect replacing the Goldstone mode by $\lambda$.

At this point we should note that by substituting (\ref{puregauge}) in $F_{ij}^B$ one gets
\be
F_{ij}^B=[\d_i Q_j-\d_j Q_i+i[Q_i,Q_j],\lambda]=[F_{ij}-i[B_i,B_j],\lambda]\label{fcond}
\ee
instead of zero. But, as we will see below eq. (\ref{ecuatia}), 
$\lambda$ and $B_i$ are of order $\mathcal O(\tfrac{1}{\tilde v})$ because Higgsing implies a term $\tilde v^2 B_iB^i$. 
This in turn means that $F^B_{ij}$ is automatically of  order $\mathcal O(\tfrac{1}{\tilde v})$ as required by 
(\ref{bconstraints}), so imposing (\ref{puregauge}) would at first seem redundant.
Yet for a purely bosonic background $F^B_{ij}$ is zero to better than $\mathcal O(\tfrac{1}{\tilde v})$ accuracy and 
(\ref{puregauge}) is needed. Since $B_i\sim  \mathcal O(\tfrac{1}{\tilde v})\rightarrow 0$, we have from (\ref{fdefs})
\be
F_{ij}\simeq \d_i Q_j -\d_j Q_i+i[Q_i,Q_j],\label{fym}
\ee
as required for a Yang-Mills theory. Then we want 
\be
|[F_{ij},\lambda]|\ll |\lambda|\;,\label{restriction}
\ee
which can be achieved in two different ways: Firstly through a restriction on the fields, by $F_{ij}$ and $\lambda$ or  $Q_i$ and $B_i$ belonging to commuting subgroups of $\SU(N)$. Secondly, we can consider that the $\U(1)$ (commuting) component of $\lambda$, $\lambda_0$, is much larger than the $\SU(N)$ components $\lambda_a$. This latter possibility has a nice physical interpretation, as we shall see.

Returning to the calculation,  the covariant derivative on a scalar $\tilde C$ becomes in terms of $\lambda$
\be
D_i \tilde C=\d_i \tilde C +i[Q_i,\tilde C]+i\{B_i,\tilde C\}=\d_i \tilde C+i[Q_i,\tilde C]+i\{D_i \lambda, \tilde C\}\equiv \tilde D_i \tilde C+i\{\d_i\lambda,\tilde C\}\;,\label{ecuatia}
\ee
with the same action for the derivative on the fermions. 

Now  consider the $\SU(N)\subset\U(N)$ part. We define $\tilde \lambda \equiv\tilde v \lambda=\tilde \lambda_a T^a$, obtaining in the large $\tilde v$ limit
\bea
&&|D_i\tilde C^{I'}|^2\rightarrow |\tilde D_i\tilde C^{I'}|^2\cr
&&\Tr |D_i\tilde C^1|^2\rightarrow \Tr |D_i\sigma|^2+(\d_i\rho_0)^2+|D_i(\rho_a+2\tilde\lambda_a)|^2\;,
\eea
while the fermionic kinetic term becomes just
\be 
-i\Tr\Big(\tilde \psi^{I\dagger}\gamma^i D_i \tilde \psi_I\Big)\rightarrow -i\Tr\Big(\tilde \psi^{I\dagger}\gamma^i \tilde D_i \tilde \psi_I\Big)\;.
\ee

Therefore as in the usual Higgs mechanism $D_i \tilde \lambda_a$ comes only in combination with $D_i \rho_a$, where $\rho_a$ is the Goldstone boson that does not appear in the scalar potential. This is how $\lambda_a$ replaces the original nonabelian Goldstone boson.

For the part of $\lambda$ which is in the $\U(1)$ centre of $\U(N)$, $\lambda_0$, the kinetic term for the scalars becomes
\be\label{lambda0}
|D_i \tilde C^I|^2=|\tilde D_i(\tilde C^Ie^{2i\lambda_0})|^2\;.
\ee
Upon taking $k$ and $\tilde v$ large the theory undergoes an effective compactification, according to the orbifold picture of \cite{Distler:2008mk}. This corresponds to the vanishing of a scalar trace degree of freedom, which in this case is $\sigma_0$.\footnote{For a precise treatment of the $\U(1)$   factors in the Higgsing of the ABJM theory see \cite{Chu:2010fk}.} The identification $\tilde C^I\sim e^{-2i\lambda_0}\tilde C^I$ signals that this degree of freedom is now carried by $\lambda_0$. Hence we also see that when solving the restriction (\ref{restriction}) by $\lambda_0\gg \lambda_a$, the interpretation is that the relative separations in the compactified direction, $\lambda_a$, are much smaller than the center of mass position $\lambda_0$.

Putting everything together, the final action is the action of a D1-brane in 9d flat space, with 7 nontrivial transverse scalars and one Goldstone boson ($\lambda$), encoding information about the 10th (compact) dimension.

\subsubsection{Higgsed action at large $\tilde v$ and D1-brane in 10d}

In getting the D1 action in 9d we eliminated the auxiliary scalars $\phi_i$, or equivalently $\tilde B_y$ and $\tilde Q_y$, via their equations of motion. However, one can easily observe that upon performing a partial integration $B_i$ also appears as an auxiliary field in the dimensionally reduced action (\ref{redaction}). In this section we will examine the consequences of eliminating $\tilde B_y$ and $B_i$ from the action instead of $\phi_i$.

We begin with the large-$\tilde v $-limit  expression coming from  (\ref{sphi}) plus the mass term for $B_i$,
\be
S_{B_i}=\int d^2 x\Big[\frac{k}{4\pi}\epsilon^{ij}\tilde Q_yF_{ij}^B-4\tilde v^2 B_iB^i+{\cal O}(\tilde v)\Big]\;.
\ee
Solving for the $\tilde B_y$ and $B_i$ auxiliary fields, we obtain
\bea
\tilde B_y&=&\frac{\sqrt{2}R}{4g\tilde v^2}\epsilon^{ij}F_{ij}+\frac{R}{4\tilde v^2}(\tilde \psi_{I\a}(\tilde \psi^{I\dagger}\gamma_3)^\a-
\tilde \psi^{I\dagger}\gamma_3\tilde \psi_I)\cr
B_i&=&\frac{k}{16\pi \tilde v^2}\epsilon_{ij}D^j\tilde Q_y\;.\label{bs}
\eea
Substituting back in the action $S_\phi+S_{B_i}$, we arrive at
\bea
S_{aux}&=&\int d^2 x\Big[-\frac{1}{4g_{YM}^2}F_{ij}F^{ij}+\frac{\sqrt{2}}{8g\tilde v^2}\epsilon^{ij}F_{ij}(\tilde \psi^{I\a}(\tilde 
\psi^{I\dagger}\gamma_3)^\a-\tilde \psi^{I\dagger}\gamma_3\tilde \psi_I)\cr
&&+\frac{1}{16\tilde v^2}(\tilde \psi^{I\a}(\tilde \psi^{I\dagger}\gamma_3)^\a-
\tilde \psi^{I\dagger}\gamma_3\tilde \psi_I)^2-\frac{k^2}{(8\pi \tilde v)^2}(D_i\tilde Q_y)^2\Big]+{\cal O}(\tfrac{1}{\tilde v})\cr
&\rightarrow&\int d^2 x\Big[-\frac{1}{4g_{YM}^2}F_{ij}F^{ij}-\frac{1}{8R^2g_{YM}^2}(D_i\tilde Q_y)^2\Big]+{\cal O}(\tfrac{1}{\tilde v})\label{saux}\;.
\eea
We note from (\ref{bs}) that $B_i$ is of order $\mathcal O(\tfrac{1}{\tilde v})$, as also mentioned in the previous subsection. Hence, $F_{ij}$ in (\ref{fdefs}) reduces to the usual Yang-Mills form (\ref{fym}).  One already sees that $\tilde Q_y$ plays the role of an extra dynamical scalar similar to $\lambda$ from the previous analysis, although unlike that case it will combine with the others to make the scalar potential of the D1-brane in 10d. For that, we are missing the terms $g_{YM}^2[\tilde X^1,\tilde X^{A'}]^2$.

Indeed, in the second line of (\ref{sphi}) one has a term 
\be
\frac{1}{4R^2}[\tilde C^I,\tilde Q_y][\tilde C^\dagger_I,\tilde Q_y]\;,
\ee
which gives the missing terms upon the identification $\tilde X_a^1= \frac{1}{2Rg_{YM}}\tilde Q_{y\; a}$. The latter also gives the correct scalar field normalisation in (\ref{saux}). Similarly, there is a missing term in the fermionic potential which is provided by the term 
\be
\frac{1}{2R} \tilde\psi^{I \dagger}\gamma_3[\tilde\psi_I,\tilde Q_y]
\ee
in (\ref{sphi}).

All in all, in the large $\tilde v$ limit and in terms of $\tilde Q_y, Q_i$, we obtain the D1-brane action in 10d flat space.

\subsubsection{Higgsed action and T-duality}\label{Tduality}

The fact that we can obtain the low-energy D1-brane action in 9d or 10d, depending on what auxiliary fields we choose to eliminate from the `master' action (\ref{redaction}), might seem strange at first. However, we will see shortly that one can switch between them using the Buscher rules \cite{Buscher:1987sk}, \ie a field theoretic version of T-duality. 

We should clarify that in our setup the M-theory direction has already been compactified on a circle of radius $R$. That led to a worldvolume reduction of the M2-brane theory. On the other hand, the T-duality we are referring to here in the $\tilde v\to\infty$ limit is on an additional compact dimension of radius $R_{10}$, which is transverse to the branes and hence involves a worldvolume scalar field. As a result, one of the 10 dimensions of string theory is compactified on a very small/very large radius.

We start with a brief review of the rules for a string worldsheet in some nontrivial background. Taking
\be
S=\int d^2 \sigma \sqrt{\gamma}\gamma^{\mu\nu}g_{ab}\d_\mu x^a\d_\nu x^b\label{original}
\ee
one writes it in a first order form as
\be
S=\int d^2\sigma \Big[\sqrt{\gamma}\gamma^{\mu\nu}(g_{00}V_\mu V_\nu+2g_{0i}V_\mu \d_\nu x^i+g_{ij}\d_\mu x^i \d_\nu x^j)+2\epsilon^{\mu\nu}\hat x^0
\d_\mu V_\nu\Big]\label{master}\;.
\ee
Then by varying with respect to $\hat x^0$ we get $\epsilon^{\mu\nu}\d_\mu V_\nu=0$, solved by $V_\mu=\d_\mu x^0$, which when plugged in 
(\ref{master}) gives back (\ref{original}). If we instead solve for $V_\mu=\frac{\hat g_{00}}{\sqrt{\gamma}}\epsilon_{\nu\mu}\d^\nu \hat x^0-\hat 
g_{0i}\d_\mu \hat x^i$ and substitute back into the action, we get the T-dual expression
\be
S=\int d^2 \sigma \Big[\sqrt{\gamma}\gamma^{\mu\nu}\hat g_{ab}\d_\mu \hat x^a\d_\nu \hat x^b+\epsilon^{\mu\nu}\hat h_{ab}\d_\mu \hat x^a\d_\nu \hat x^b
\Big]\;,
\ee
where the background fields are the T-dual ones:
\be
\hat g_{00}=\frac{1}{g_{00}};\;\;\;
\hat g_{ij}=g_{ij}-\frac{g_{0i}g_{0j}}{g_{00}};\;\;\;
\hat h_{0i}=\frac{g_{0i}}{g_{00}}\;,
\ee
with the remaining components of $\hat h_{ab}$ zero and $\hat x^i = x^i$.

In order to compare the above with our case, we concentrate on the relevant terms in the two first order `master' actions\footnote{The difference in the  overall sign is due to different conventions.}
\bea
S_{D1}&=&-\int[4\tilde v^2B_i^2-\frac{k}{4\pi}\epsilon^{ij}\tilde Q_y F_{ij}^B]\nn\\
S_{Buscher}&=&\int [g_{00}V_\mu V^\mu+2\epsilon^{\mu\nu}\hat x^0\d_\mu V_\nu]\;,
\eea
which leads  to the identifications
\bea
V_\mu&\leftrightarrow& B_i\cr
g_{00}&\leftrightarrow & 4 \tilde v^2\cr
\hat x^0&\leftrightarrow & -\frac{k}{4\pi}\tilde Q_y\cr
x^0&\leftrightarrow & \lambda\;.\label{mapul}
\eea
The T-dual second order forms are compared in a similar manner:
\bea
g_{00}\d_\mu x^0\d^\mu x^0&\leftrightarrow & 4\tilde v^2 (\d_i \lambda)^2\cr
\frac{1}{g_{00}}\d_\mu \hat x^0\d^\mu \hat x^0&\leftrightarrow& \frac{1}{4\tilde v^2}\Big[D_i\Big(-\frac{k\tilde Q_y}{4\pi}\Big)\Big]^2\;.
\eea

We therefore see that the $R_{10}\rightarrow \tfrac{1}{R_{10}}$ T-duality  in our case becomes $2\tilde v\rightarrow \tfrac{1}{2\tilde v }$. This indicates that while in the first formulation
\be
\lambda\sim \frac{1}{\tilde v}, 
\ee
\ie $\lambda$  takes values on a circle of small radius $\propto \tfrac{1}{\tilde v}$, as expected for a D1-brane in 9d, in the T-dual formulation
\be
\frac{k\tilde Q_y}{4\pi}\sim \tilde v
\ee
and the T-dual field takes values on a circle of large radius $\propto \tilde v$, with the corresponding direction  decompactified as expected for a D1-brane in 10d.

One is perhaps more familiar with T-duality exchanging the momentum $p=\frac{n}{R_{10}}$ with winding $w=\frac{mR_{10}}{\a'}$, or $n$ with $m$ in the 
expansion of the physical compact direction $R_{10}x^0=X^0=\frac{n\a'}{R_{10}}\tau+mR_{10}\sigma+\ldots$ . But as is well-known, 
in the Buscher form of T-duality this arises because the two dual coordinates $x^0$ and $\hat x^0$ are solutions to the same master $V_\mu$ as
\be
V_\mu=\d_\mu x_0=\frac{\hat g_{00}}{\sqrt{\gamma}}\epsilon_{\nu\mu} \d^\nu \hat x^0-\hat g_{0i}\d_\mu \hat x^i\;.
\ee
Taking the particular case $g_{0i}=0$ and the conformal gauge $\gamma_{\mu\nu}=\delta_{\mu\nu}$, one gets for the physical coordinates $\d_\mu (R_{10}x^0)=\epsilon_{\mu\nu}\d^\nu (\frac{\a'}{R_{10}} \hat x^0)$, or $\d_\tau (X^0)=\d_\sigma({X^0}')$, $\d_\sigma(X^0)=\d_\tau({X^0}')$, \ie exactly exchanging $p$ with $w$. In our case, we can use the map (\ref{mapul}) to obtain the exact same momentum $\leftrightarrow$ winding exchange in the limit where we only keep the compact abelian scalar from the action. Of course, in the full theory it is not clear how the T-duality acts on full states.

\subsection{The $N=1$ case}\label{N=1}

We next turn to the study of the abelian  case, which we will further interpret in the following section. By setting $N=1$ in the ABJM action, one obtains a theory for a single supermembrane on $\mathbb C^4/\mathbb Z_k$ \cite{Aharony:2008ug,Sasaki:2009ij}
\be\label{single}
S^{N=1}_{\mathrm{ABJM}}=\int d^3 x\left[\frac{k}{4\pi}\epsilon^{\mu\nu\lambda}\left(A_\mu^{(1)}\d_\nu A_\lambda^{(1)}-A_\mu^{(2)}\d_\nu A_\lambda^{(2)}\right) -i\psi^{I\dagger}\gamma^\mu D_\mu \psi_I-D_\mu C^\dagger_I D^\mu C^I\right]\;,
\ee
where
\be
D_\mu C^I=\d_\mu C^I-i(A_\mu^{(1)}-A^{(2)}_\mu)C^I\;.
\ee

By applying the same dimensional reduction procedure as for the general nonabelian case  we obtain the 2d action
\bea
S&=&\int d^2 x\Big[-D_i\tilde C_I^\dagger D^i\tilde C^I-i\tilde \psi ^{I\dagger}\gamma^iD_i\tilde \psi_I\Big]+S_\phi\label{Dactionzero}\\
S_\phi&=&\int d^2 x \Big[\frac{\sqrt{2}}{2g R}\epsilon^{ij}(\phi_1 F_{ij}^{(1)}-\phi_2 F_{ij}^{(2)})
-R^{-2}(\tilde C^I\tilde C^\dagger_I)(\phi_1-\phi_2)^2\cr
&&\qquad\qquad \qquad \qquad \qquad  \qquad -R^{-1}(\phi_1-\phi_2)\tilde \psi^{I\dagger}\gamma_3\psi_I\Big]\;.
\eea
Then varying with respect to $\phi_1,\phi_2$ we get the constraints
\bea
\phi_1-\phi_2&=&\frac{R\sqrt{2}}{4g}\frac{\epsilon^{ij}F_{ij}^{(1)}}
{\tilde C^I\tilde C^\dagger_I}-\frac{R}{2\tilde C^I\tilde C^\dagger_I}
(\tilde \psi^{I\dagger}\gamma_3\psi_I)\cr
\epsilon^{ij}F_{ij}^{(2)}&=&\epsilon^{ij}F_{ij}^{(1)}\label{constrN1}
\eea
and once again  $\phi_1-\phi_2$  as well as $F^{(2)}_{ij}$ can be solved in terms of other fields.
Substituting back in $S_\phi$:
\bea
S_\phi&=&\int d^2 x\Big[-\frac{1}{4g^2 \tilde C^I\tilde C^\dagger_I}(F_{ij}^{(1)})^2
+\frac{1}{4\tilde C^I\tilde C^\dagger_I}(\tilde \psi^{I\dagger}
\gamma_3\tilde \psi_I)^2-\frac{\sqrt{2}}{4g \tilde C^I\tilde C^\dagger_I}\epsilon^{ij}F_{ij}^{(1)}\tilde \psi^{I\dagger}\gamma_3\tilde \psi\Big]
\label{Daction}\nn\\
&=&\int d^2 x \frac{\tilde C^I\tilde C^\dagger_I}{R^2}\Big[\frac{R\sqrt{2}}{4g}\frac{\epsilon^{ij}F_{ij}^{(1)}}
{\tilde C^I\tilde C^\dagger_I}-\frac{R}{2\tilde C^I\tilde C^\dagger_I}(\tilde \psi^{I\dagger}\gamma_3\psi_I)\Big]^2\label{rewrite}\;.
\eea

The equation of motion for $A_i^{(1)}$  gives 
\be
\d_i\Big[\frac{R\sqrt{2}}{4g}
\frac{\epsilon^{ij}F_{ij}^{(1)}}{\tilde C^I\tilde C^\dagger_I}-\frac{R}{2\tilde C^I\tilde C^\dagger_I}
(\tilde \psi^{I\dagger}\gamma_3\psi_I)\Big]=0\label{eomai}\;,
\ee
\ie the bracket is a constant, with $S_\phi$  proportional to  the square of this bracket. Note that due to the constraints (\ref{constrN1}) we can choose a gauge in which $A^{(1)}_i=A^{(2)}_i$,  such that the covariant derivative reduces to the partial derivative.

\subsection{Supersymmetry}

The action has 12 supersymmetries for any value of $N$, found by dimensional reduction of the ABJM supersymmetries.
The supersymmetry rules in 3d are 
\bea
\delta C^I&=&i\omega^{IJ}\psi_J\cr
\delta C^\dagger_I&=&i\psi^{\dagger J}\omega_{IJ}\cr
\delta\psi_I&=& -\gamma_\mu \omega_{IJ}D_\mu C^J+\frac{2\pi}{k}(-\omega_{IJ}(C^KC^\dagger_KC^J-C^JC^\dagger_KC^K)+2\omega_{KL}C^K C^\dagger_I C^L)\cr
\delta \psi^{I\dagger}&=&D_\mu C^\dagger_J\gamma_\mu \omega^{IJ}+\frac{2\pi}{k}(-(C^\dagger_JC^KC^\dagger_K-C^\dagger_KC^KC^\dagger_J)+2C^\dagger_LC^I
C^\dagger_K\omega^{KL})\cr
\delta A^{(1)}_\mu&=&-\frac{\pi}{k}(C^I\psi^{J\dagger}\gamma_\mu \omega^{IJ}+\omega^{IJ}\gamma_\mu \psi_I C^\dagger_J)\cr
\delta A^{(2)}_\mu &=& \frac{\pi}{k}(\psi^{I\dagger}C^J\gamma_\mu \omega_{IJ}+\omega^{IJ}\gamma_\mu C^\dagger_I\psi_j)\;.
\eea
Dimensionally reducing, we obtain
\bea
\delta\tilde C^I&=&i\omega^{IJ}\tilde\psi_J\cr
\delta\tilde \psi_I&=&-\gamma_i\omega_{IJ}D_i \tilde C^J-\frac{i}{R}\gamma_3\omega_{IJ}(\tilde C^J\phi_2-\phi_1\tilde C^J)\cr
&&+\frac{g }{\sqrt{2}}[-\omega_{IJ}(\tilde C^K\tilde C^\dagger_K\tilde C^J-\tilde C^J\tilde C^\dagger_K\tilde C^K)
+2\omega_{KL}\tilde C^K\tilde C^\dagger_I\tilde C^L]\cr
\delta A^{(1)}_i&=&-\frac{g}{2\sqrt{2}}(\tilde C^I\tilde \psi^{J\dagger}\gamma_i\omega_{IJ}+\omega^{IJ}\gamma_i\tilde \psi_I\tilde C^\dagger_J)\cr
\delta A^{(2)}_i&=&+\frac{g}{2\sqrt{2}}(\tilde \psi^I\tilde C^J\gamma_i\omega_{IJ}+\omega^{IJ}\gamma_i\tilde C_I\tilde \psi_J)\cr
\delta \phi_1&=&-\frac{Rg}{2\sqrt{2}}(\tilde C^I\tilde \psi^{J\dagger}\gamma_3\omega_{IJ}+\omega^{IJ}\gamma_3\tilde \psi_I\tilde C^\dagger_J)\cr
\delta \phi_2&=&+\frac{Rg}{2\sqrt{2}}(\tilde \psi^I\tilde C^J\gamma_3\omega_{IJ}+\omega^{IJ}\gamma_3\tilde C_I\tilde \psi_J)\;.
\eea
Restricting to $N=1$
\bea
\delta\tilde C^I&=&i\omega^{IJ}\tilde \psi_J\cr
\delta\tilde \psi_I&=&-\frac{1}{R}\gamma_i \omega_{IJ}D_i\tilde C^J+i\gamma_3\omega_{IJ}\tilde C^J(\phi_1-\phi_2)\cr
\delta A^{(1)}_i=\delta  A^{(2)}_i&=&-\frac{g}{2\sqrt{2}}(\tilde C^I\tilde \psi^{J\dagger}\gamma_i\omega_{IJ}+\omega^{IJ}\gamma_i\tilde \psi_I\tilde 
C^\dagger_J)\cr
\delta\phi_1=\delta\phi_2&=&-\frac{Rg}{2\sqrt{2}}(\tilde C^I\tilde \psi^{J\dagger}\gamma_3\omega_{IJ}+\omega^{IJ}\gamma_3\tilde \psi_I\tilde 
C^\dagger_J)\;.
\eea
However, if we were to solve for $\phi_1,\phi_2$, we would get 
\bea
\delta\tilde C^I&=&i\omega^{IJ}\tilde \psi_J\cr
\delta\tilde \psi_I&=&-\gamma_i \omega_{IJ}D_i\tilde C^J+i\gamma_3\omega_{IJ}\tilde C^J
\Big[\frac{\sqrt{2}}{4g}\frac{\epsilon^{ij}F_{ij}^{(1)}}{\tilde C^I\tilde C^\dagger_I}-\frac{1}{2\tilde C^I\tilde C^\dagger_I}
(\tilde \psi^{I\dagger}\gamma_3\psi_I)\Big]\cr
\delta A^{(1)}_i&=&-\frac{g}{2\sqrt{2}}(\tilde C^I\tilde \psi^{J\dagger}\gamma_i\omega_{IJ}+\omega^{IJ}\gamma_i\tilde \psi_I\tilde 
C^\dagger_J)\;.
\eea
This is only an {\it on-shell} supersymmetry of the action with $\phi_i$ eliminated (specifically, on-$A^{(1)}_i$-shell, \ie using (\ref{eomai})), since for instance by varying the 
$\tilde C^I$ in (\ref{rewrite}) we get  unique terms involving $(\tilde C^I\tilde C^\dagger_I)^{-2}$, of the type
\be
 -\frac{1}{R^2}(\tilde C^I\delta\tilde C^\dagger_I)\Big[\frac{R\sqrt{2}}{4g \tilde v }\frac{\epsilon^{ij}F_{ij}^{(1)}}{\tilde C^I\tilde C^\dagger_I}
-\frac{R}{2\tilde C^I\tilde C^\dagger_I}(\tilde \psi^{I\dagger}\gamma_3\psi_I)\Big]^2\;.
\ee
However,  when using the $A^{(1)}_i$ equation of motion, the above becomes of similar type to other more conventional terms.

\section{Interpretational points}

We now move towards interpreting the dimensionally reduced ABJM action  obtained in the previous section. This turns out to hide certain  subtleties and involves an order of limits.

\subsection{From ABJM to BST formulation for the membrane}

Let us begin with the case of the single M2-brane in the Green-Schwarz-type supermem\-brane description of Bergshoeff-Sezgin-Townsend (BST) \cite{Bergshoeff:1987cm}. In a general supergravity background this is given by
\be
S=\int d^3 x \Big[\frac{1}{2}\sqrt{-g}g^{\mu\nu}E_\mu^AE_\nu^B\eta_{AB}+\frac{1}{2}\epsilon^{\mu\nu\lambda}E_\mu^AE_\nu^BE_\lambda^C C_{ABC}-\frac{1}{2}\sqrt{-g}\Big]\;,
\ee
where 
\be
E_\mu^A=(\d_\mu Z^M)E_M^A\;.
\ee
In the above the $Z^M$ are superspace coordinates while $E_M^A$ is the supervielbein, such that 
\be
g_{\mu\nu}=E_\mu^AE_\nu^B\eta_{AB}\;,
\ee
where $\mu= 0,1,2$ are  worldvolume while $A=0,...,10$ spacetime indices.  The superfields satisfy the 11d supergravity constraints. The bosonic degrees of freedom are $X^A$, which in  static gauge reduce to the  $C^I$ scalars, and involve  no gauge fields.

At the same time, for the $N=1$ ABJM action of (\ref{single}) and for $k=1$, one once again  expects to obtain the static gauge action for a single supermembrane in flat space. Compared to the BST approach, this is a formulation that does involve auxiliary gauge fields.  

Naturally, the formulations with and without gauge fields should be equivalent and this can be established as follows:\footnote{A version of this procedure embedded in the nonabelian theory also 
appears in \cite{Lambert:2010ji}.} Since the gauge fields are abelian, we can rewrite (\ref{single}) in terms of their sum and difference
\bea
Q_\mu &=& (A^{(1)}_\mu + A^{(2)}_\mu)\nn\\
B_\mu &=& (A^{(1)}_\mu - A^{(2)}_\mu)\;,
\eea
obtaining
\be
S^{N=1}_{\mathrm{ABJM}}=\int d^3 x\left[\frac{k}{4\pi}\epsilon^{\mu\nu\lambda} B_\mu\d_\nu Q_\lambda -i\psi^{I\dagger}\gamma^\mu D_\mu \psi_I-D_\mu C^\dagger_I D^\mu C^I\right]\;.
\ee
We next define the field strength $H_{\mu\nu } = \pd_\mu Q_\nu - \pd_\nu Q_\mu$ and treat it as an independent field. This is achieved by introducing a Lagrange 
multiplier that imposes the Bianchi identity on $H_{\mu\nu}$ through the equations of motion for the `dual photon' $\sigma$:
\be\label{dualphoton}
S^{N=1}_{\mathrm{ABJM}}=\int d^3 x\left[\frac{k}{8\pi}\epsilon^{\mu\nu\lambda} B_\mu H_{\nu\lambda} +\frac{1}{8\pi} \sigma\epsilon^{\mu\nu\lambda} \partial_{\mu}H_{\nu\lambda} -i\psi^{I\dagger}\gamma^\mu D_\mu \psi_I-D_\mu C^\dagger_I D^\mu C^I\right]\;.
\ee
Integrating this new term by parts we find
\be
S^{N=1}_{\mathrm{ABJM}}=\int d^3 x\left[\frac{k}{8\pi}\epsilon^{\mu\nu\lambda} B_\mu H_{\nu\lambda} -\frac{1}{8\pi}\epsilon^{\mu\nu\lambda} (\pd_\mu \sigma) H_{\nu\lambda} -i\psi^{I\dagger}\gamma^\mu D_\mu \psi_I-D_\mu C^\dagger_I D^\mu C^I\right]\;.
\ee
It is now possible to integrate out $H_{\mu\nu}$, arriving at  the relation
\begin{equation}
B_\mu = \frac{1}{k}\partial_\mu \sigma\;,
\end{equation}
with the $\U(1)_B$ gauge transformations acting on the dual photon as
\begin{equation}\label{fix}
\sigma \to \sigma +  {k}\theta\ .
\end{equation}

Accordingly the covariant derivatives become
\be
D_\mu C^I = \pd_\mu C^I - i (A^{(1)} - A^{(2)}) C^I = \pd_\mu C^I - i B_\mu C^I = \pd_\mu C^I - \frac{i}{k} \pd_\mu \sigma C^I
\ee
and the action can be rewritten in terms of a new set of matter fields
\be
\hat C^I = e^{-\frac{i \sigma}{k}} C^I\qquad\textrm{and}\qquad \hat \psi_I = e^{-\frac{i \sigma}{k}} \psi_I
\ee
resulting in
\be\label{singleM2}
S^{N=1}_{\mathrm{ABJM}}=\int d^3 x\left[ -i\hat\psi^{I\dagger}\gamma^\mu \pd_\mu \hat\psi_I-\pd_\mu \hat C^\dagger_I \pd^\mu \hat C^I\right]\;.
\ee
In this manner the auxiliary gauge fields have been eliminated from the action. We note that as $\sigma$ is dual to the $\U(1)_B$ gauge field, it is a compact scalar with an associated periodic shift-symmetry.

However the above still transform under the $\U(1)_B$ gauge transformations, which we will next gauge-fix. In order to do so, we need to determine the periodicity of $\sigma$ which follows from a quantisation condition on the flux $H$.  We have already defined $H = F^{(1)} + F^{(2)}$. By imposing the standard Dirac quantisation condition on the original gauge fields $\int d F^{(1,2)}\in 2\pi \mathbb Z$, we get
\begin{equation}\label{Hflux}
\int \frac{1}{2}\epsilon^{\mu\nu\lambda}\partial_\mu H_{\nu\lambda}= 
\int d H = \int d F^{(1)} + \int d F^{(2)} \in 4\pi \mathbb{Z}
\end{equation}
and by plugging this into (\ref{dualphoton})  and requiring that the path integral remains invariant under periodic shifts of $\sigma$, we determine that the latter has period $2\pi$ \cite{Lambert:2010ji}. This can be then used to gauge-fix the $\U(1)_B$ symmetry through (\ref{fix}) and set $\sigma=0\ {\rm mod }\ 2\pi$. However, this periodicity imposes an additional identification on the $\U(1)$-invariant fields $\hat C^I, \hat \psi_I$
\begin{equation}\label{U(1)b}
\hat Z^A \cong e^{-\frac{2\pi  i}{k}}\hat Z^A\qquad \textrm{and} \qquad \hat \psi_A \cong
e^{-\frac{2\pi  i}{k}}\hat \psi_A\ .
\end{equation}
Therefore, we have obtained that (\ref{singleM2}) along with the above identification is nothing but the dynamical part for the action of a single BST M2-brane propagating on a $\mathbb C^4/\mathbb Z_k$ background. For the case of $k=1$, one recovers the action for a single membrane in flat space.

\subsection{From the reduced action to a Green-Schwarz string action}\label{stringequiv}

Having established the relationship between the $N=1$ ABJM and BST actions, we now repeat the argument in the dimensionally reduced theory, which will exhibit some subtle points. 
As before, the reduced coordinate is denoted by $y$ and $i=0,1$ indicate the 2d coordinates.
Dimensionally reducing we get for $B_\mu$ and $Q_\mu$\footnote{Note that in this subsection we define $B_i,Q_i$ without 
a prefactor of $\tfrac{1}{2}$.}
\bea
B_y=\frac{1}{R}(\phi_1-\phi_2)\;, && Q_y=\frac{1}{R}(\phi_1+\phi_2)\nn\\
B_i=A_i^{(1)}-A_i^{(2)}\;,&& Q_i=A_i^{(1)}+A_i^{(2)}\;.
\eea
Expressing the abelian CS piece of (\ref{single}) in terms of $B$ and $Q$, we get
\be
S_{CS}= \frac{k}{4\pi}\int d^3 x\epsilon^{\mu\nu\lambda}B_\mu\d_\nu Q_\lambda\rightarrow
\frac{kR}{4\pi}\int d^2x[B_y\epsilon^{ij}\d_i Q_j+\epsilon^{ij}B_i\d_j Q_y]\label{CSorig}\;.
\ee
We define $H_{ij}\equiv 2\d_{[i} Q_{j]}$ but cannot treat it as an independent field, since it is not possible to introduce a Lagrange multiplier that imposes its
Bianchi identity. On the other hand, we can obtain an independent field $H_{jy}$ by first defining $H_{jy}=\d_j Q_y$ and then introducing it with a Lagrange multiplier $\tilde \sigma$. The equivalent CS action in 2d is then
\be
S_{CS}=\frac{R}{8\pi}\int d^2 x\Big[k B_y\epsilon^{ij}H_{ij}+k\epsilon^{ij}B_i H_{jy}+\tilde\sigma\epsilon^{ij}\d_i H_{jy}\Big]\label{CSnew}\;.
\ee

The rest of the dimensionally reduced action is easily obtained from (\ref{redaction}) in the abelian limit
\be
\int d^2 x [-D_i \tilde C^\dagger_I D^i\tilde C^I-i\tilde \psi^{I\dagger}\gamma^iD_i\tilde \psi_I- (B_y)^2\tilde C^I\tilde C^\dagger_I
-B_y\tilde \psi^{I\dagger}\gamma_3\psi_I]\label{restorig}
\ee
where the last two terms come from $D_y\sim-iB_y$ terms.

Now, if in the dimensionally reduced gauge action one integrates out $\phi_1,\phi_2$, that is $B_y$ and $Q_y$ in (\ref{CSorig}) plus (\ref{restorig}),
one obtains (\ref{Dactionzero}) with (\ref{Daction}). But if we instead  go to (\ref{CSnew}) and eliminate $H_{jy}$ we get
\be
B_i=\frac{1}{k}\d_i \sigma
\ee
 and then, as in the previous section, 
\be
D_i \tilde C^I\rightarrow \d_i\tilde C^I-\frac{i}{k}\d_i \sigma \tilde C^I\;.
\ee
By eliminating $Q_j$
\be
\d_i B_y=0\Rightarrow B_y=c\label{B0}\;,
\ee
where $c$ some constant. For the reader who might worry that $B_y=\tfrac{1}{R}(\phi_1-\phi_2)=c$ seems to contradict (\ref{constrN1}), we should note that it is  consistent with (\ref{B0}) on-shell, since the equation of motion for $A^{(1)}$ implies 
\be
\frac{\sqrt{2}}{4g \tilde v}\epsilon^{ij}F_{ij}^{(1)}-\tilde \psi^{I\dagger}\gamma_3\tilde \psi_I=c\, \tilde C^I\tilde C^\dagger_I\label{constraintnew}\;,
\ee
which is the same as (\ref{eomai}).

Finally with the redefinitions
\be
\hat{ \tilde{C}}^I=e^{-\frac{i\sigma}{k}}\tilde C^I;\;\;\;
\hat{ \tilde{\psi}}_I=e^{-\frac{i\sigma}{k}}\tilde \psi_I
\ee
we arrive at the action
\be
S=\int d^2 x \Big[-\d_i \hat{\tilde{C} }^I\d^i \hat{\tilde{C}}^\dagger_I-i\hat{\tilde{\psi} }^{I\dagger}\gamma^i\d_i\hat{\tilde{\psi} }_I
-c^2\hat{\tilde{C} }^I\hat{\tilde {C} }^\dagger_I-c\hat{\tilde{\psi} }^{I\dagger}\gamma_3\hat{\tilde{\psi} }_I\Big]\label{GSstringaction}\;,
\ee
which is a Green-Schwarz-type action plus some arbitrary mass terms. These can be put to zero by choosing $c=0$, since $c$ is an arbitrary constant at this point. We will come back to the interpretation of this constant soon.

\subsection{Interpretation of the 2d action}

So far we have only discussed the set of algebraic steps that relate to the reduction process. We now turn our attention to assigning an interpretation to the resulting action.

\subsubsection{Intuition from String/M-theory}

As described in the introduction, the natural expectation is that reducing an action for M2-branes on a circle should lead to fundamental strings. We indeed discussed in Section \ref{stringequiv} how the abelian, dimensionally-reduced ABJM action can be transformed to a Green-Schwarz action on an orbifold background. On the other hand, the presence of a single gauge field on the 2d worldvolume before the transformation would na\"ively suggest that it describes a D1-brane.

%F- and D-strings are related by an S-duality, which means that they should be valid in different limits of the $g_s$ parameter space. In Einstein frame the D1-brane tension goes like $g_s^{-1/2}$, while for the fundamental string like $ g_s^{1/2}$. Hence, for weak string coupling the lightest scale in the theory is the fundamental string tension and one has a theory of strings, while at strong string coupling the lightest scale is the D-string tension and one expects a theory of weakly coupled D1-branes. 

To further understand this, consider the case of M-theory on $T^2$. The compactification of a single M2-brane on the 2 circles (one transverse, $R_2$, one parallel, $R_1$) depends on the order of compactification: Indeed, if we compactify first on $R_2$, we get a D2-brane in 10d, with $g_s=\tfrac{R_2}{l_s}$. A further $R_1$ compactification must be followed by T-duality on $R_1$, to obtain a D1 in IIB, upon which  $g_s\rightarrow g_s\tfrac{l_s}{R_1}=\tfrac{R_2}{R_1}$. On the other hand, by first compactifying on $R_1$ we get an F1 in 10d but with $g_s=\tfrac{R_1}{l_s}$. By further compactifying  on $R_2$ we must  perform a T-duality in order to arrive at  IIB, getting $g_s=\tfrac{R_1}{R_2}$. The two cases are related by $g_s\rightarrow \tfrac{1}{g_s}$, \ie S-duality in 9d, which takes the fundamental string to a D1-brane.

However, the string coupling enters the two actions in a different way: For D1's it appears through $(g^{D1}_{YM})^2 = \tfrac{g_s}{2 \pi \alpha'}$, in canonical normalisation, while for F1's as an additive term $\int d^2x \phi R^{(2)}\sim \ln g_s\chi$, with $\chi$ the Euler characteristic of the worldsheet. As a result one cannot make the duality precise at the level of the actions in their conventional formulations.

\subsubsection{Intuition from Higgsing}

We will gain some further insight into the physics of our action from looking at the Higgsed theory.
%Note that below we have general N, so it does not make sense to talk about N=1. 
As we have already seen, in the presence of a VEV $
\langle\tilde{C}^I\rangle=\tilde{v}\delta^{I}_1 \one_{N \times N}$ the action (\ref{YMaction})  at general $N$ 
has a gauge coupling 
\be 
g\,\tilde{v}=\frac{2\pi   \sqrt{2}\tilde{v}}{kR}\equiv g^{D1}_{YM}\;.  
\ee 
Note that as far as $g^{D1}_{YM}$ is concerned, a finite coupling can be obtained by having $k$ and $\tilde v$ either be generic or both large, so one needs to look at other criteria. Furthermore, the coupling does not differentiate between a D1-brane in 9 or 10 dimensions. 

By  first considering the case of {\it large $\tilde v$ and large $k$}, we indeed obtained a D1-brane either in 9 or 10 dimensions, depending on the fields that were integrated out: By choosing $\phi_1$ and $\phi_2$, \ie $\tilde B_y$ and $\tilde Q_y$, the compact scalar was eaten by the massive vector $B_i$ through a version of the ordinary Higgs mechanism, which was in turn replaced by a new scalar $\lambda$.  For the T-dual version of the dimensionally reduced ABJM theory, $Q_i$ and the $\tilde Q_y=\phi_1 + \phi_2$ combination were traded for the scalar corresponding to the compact dimension and became dynamical in the Higgsed theory, in the spirit of \cite{Mukhi:2008ux}.

One needs to keep in mind that the compactification picture is valid at worldvolume energies $E\ll1/R$ and that a prospective D1-brane description would be weakly coupled if 
\be 
g_{eff}=\frac{g^{D1}_{YM}}{E}=2\pi \sqrt{2}\frac{\tilde v}{k}\frac{1}{ER}\ll 1\;.  
\ee 
We see that this is only possible for small $\tfrac{\tilde v}{k}$, which matches with the intuition that 
we need to consider large $k$ to compactify a transverse scalar. We also need to be in 
an intermediate energy regime where
\be
 (2\pi \sqrt{2})\frac{\tilde v}{k}\frac{1}{R}\ll E\ll\frac{1}{R}\;.  
\ee 

At {\it generic $k$} the above theory is strongly coupled. At {\it generic} $\tilde v$ there are $\mathcal O(\tfrac{1}{\tilde v})$ corrections and the theory is not just SYM. However, in both these cases, one would instead expect an S-dual F1 description, at least if one eliminates $\phi_i$ as was done for $N=1$ in (\ref{Dactionzero})-(\ref{Daction}).  From this angle, the fact that the action still has a worldvolume gauge field is not crucial, since we saw in Section \ref{stringequiv} that it can be integrated out upon explicitly imposing the orbifold conditions on the matter fields to obtain a Green-Schwarz type action for the string.

\subsubsection{Reducing/T-dualising ``M2 to D2''}

The Higgsed actions can also be obtained from the worldvolume reduction or T-duality of the ABJM membrane through the mechanism of \cite{Mukhi:2008ux}, after giving a large VEV $v$ to one of the original ABJM worldvolume scalars. That is, we can consider Higgsing the theory {\em before} performing the dimensional reduction.

In terms of a geometric description the value of $k$ is interpreted as the rank of the $\mathbb C^4/\mathbb Z_k$ orbifold singularity on which the membranes are sitting. Going off to the Coulomb branch  at large $v$ and $k$ results in type IIA String Theory dynamics. The M2 to D2 reduction is manifest in a way similar to the models of `deconstruction' \cite{ArkaniHamed:2001ie}, by having a fixed and finite radius of compactification for the transverse direction. In fact, the large-$k$ dynamics are those of IIA even at the superconformal point (zero VEVs) where one has four complex scalars, as was made apparent by the analysis of \cite{Aharony:2008ug}, though of course at finite or zero VEV one probes an M-theory radius that varies between zero and a small nonzero value.

The resulting field theory action is 3d SYM with corrections of order $\mathcal O (\tfrac{1}{ v})$ and gauge coupling
\be
g^{D2}_{YM}=\frac{2\pi \sqrt{2}v}{k}=\frac{2\pi\sqrt{2}\tilde v}{k\sqrt{R}}\;,
\ee 
where the 11d VEV $v$ is related to the 10d one by $\tilde{v}=\sqrt{R}v$. On the other hand, at generic $k$ and $\tilde v$ one remains in the M2-brane regime. 

Performing a worldvolume dimensional reduction/T-duality we again recover the D1-action in 9d and 10d respectively. This is essentially because the Higgsing and reduction/T-duality operations commute. Our dimensionally reduced action is once more better thought of as one for a fundamental string, at least if the $\phi_i$ are  eliminated, as we did for the $N=1$ case in (\ref{Dactionzero})-(\ref{Daction}).

\subsubsection{Interpretation in terms of `master' action}

We conclude that our general action Eq.~(\ref{redaction}), with all the auxiliary fields, contains information for all of the IIA F-string in 10d, its T- and S-dual  IIB D-string in 10d, as well as the latter's T-dual IIA D-string in 9d (compactified D2-brane).  The IIA F-string in 10d and the IIA D-string in 9d  are related through the compactification obtained by large $k$ together with an S-duality in 9d. Therefore we can think of the action (\ref{redaction}) as a `master' action that contains information about both T- and S-duality.

Even though we have recovered a precise field theoretic realisation of T-duality in Section \ref{Tduality}, it is difficult to disentangle how the S-duality acts. It seems  that the two components needed are the transformation of Section \ref{stringequiv} and the transverse compactification through large $\tilde v$ and large $k$.  We should also note that while the discussion made heavy use of the $N=1$ case, for which more explicit formulas were available, we still expect that at general $N$ the action (\ref{redaction}) is a `master' action for T- and S-duality.

\subsubsection{Understanding the F1-string interpretation}

There remains one point that has not yet been clarified for the `master' action.  If the equivalent action (\ref{GSstringaction}) is to have  an F1 interpretation, it should describe a string moving in the orbifold background $\mathbb C^4/\mathbb Z_k$. As such, it must provide a natural explanation  for the mass terms in (\ref{GSstringaction}). Indeed, note that $c$ has mass dimension 1, so we can instead denote it by a mass $\mu$.

Consider a straight string in $\mathbb C^4$ whose endpoints span exactly a $\tfrac{2\pi}{k}$ angle from the origin. By making the $\mathbb Z_k$ identification we create a noncontractible string with winding number 1. This can however slide towards the origin by virtue of its tension, unlike  the case of usual winding in a circle direction. Taking this string to be symmetric relative to the origin, its energy will be\footnote{If the string is asymmetric relative to the origin its tension will make it symmetric since that corresponds to a minimum length for a  given  center-of mass distance to the origin.} 
\be 
E=Tr2\sin\tfrac{\pi}{k} \;,
\ee
 where $T=\tfrac{1}{2\pi\alpha'}$ is the string tension and $r$ is the radius from the origin of spacetime to the endpoints of the string, $r=|X(0)|$.  The force pulling the string towards the origin will be 
\be 
F=\frac{dE}{dr}=2T\sin\tfrac{\pi}{k}\equiv m\frac{d^2r}{dt^2}\;,
\ee
 where $m=E=2Tr\sin\tfrac{\pi}{k}$ is the mass of the string. Substituting and introducing the appropriate sign we get 
\be 
\frac{d^2   r}{dt^2}=-\frac{1}{r}=-\frac{1}{r^2}r\equiv -\mu^2 r \;.
\ee
Here we have made assumption that due to its tension the string stays straight and symmetric as it slides towards the origin. Then the position $X(\sigma)$ along the string in some Cartesian spacetime reference system varies relative  to the endpoints, \ie 
\be
\frac{\ddot X(\sigma)}{X(\sigma)}=\frac{\ddot X(0)}{X(0)}=-\mu^2\;.
\ee
This  matches the worldvolume equation of motion of the above straight string, $X'=0$, with a worldvolume mass term $\mu$, namely
\be
\ddot{X}(\sigma)=-\mu^2 X(\sigma)\;.
\ee
Since $r$ is arbitrary then so is $\mu$, as was also the case in our worldvolume analysis.  Note that we have not needed any approximation for this result. Indeed, the equations of motion for $\mu$ were $\d_i(\mu)=0$, that is $\mu$ constant {\it on the worldvolume}, and this is what we find. Of course $\mu=\tfrac{1}{r}$, with $r$ the radius at the endpoint, so $\mu$ depends on the spacetime boundary of the string but that does not contradict our constraints.

To conclude, the arbitrary scalar mass term in (\ref{GSstringaction}) is just the effect of having a string in $\mathbb C^4/\mathbb Z_k$. The fermion mass term is understood as the necessary supersymmetric extension. This  completes our understanding of the $N=1$ 2d action as an F-string in $\mathbb C^4/\mathbb Z_k$.

\section{Conclusions}\label{conclusions}

In this note we performed the dimensional reduction of the ABJM model, studying the resulting action and its physical interpretation. The reduced action includes a set of auxiliary gauge and scalar fields. Focusing on the Coulomb branch of the theory, we found that for a large VEV $\tilde v$, in the regime where $k$ was large, one obtains either the action of $N$ D1-branes in 9d (compactified D2-branes) or the T-dual action of $N$ D1-branes in 10d, depending on which combination of auxiliary fields are integrated out. The two actions were related by a field theoretic T-duality transformation, following \cite{Buscher:1987sk}.

For the special case of $N=1$, at an arbitrary VEV $\tilde v$ and at arbitrary $k$, we showed that the equivalence in 3d between the $N=1$ ABJM model and the BST action on $\mathbb C^4/\mathbb Z_k$ can be reduced to give a Green-Schwarz string moving in $\mathbb C^4/\mathbb Z_k$. This led us to propose that the dimensionally reduced action can be thought of as a `master' action encoding information about both T- and S-duality. The field theoretic realisation of S-duality in the nonabelian case remains mysterious as ever and warrants further investigation. Since at generic $k$/VEV $\tilde v$ the 9d D1-brane action obtained by eliminating $\phi_i$ is strongly-coupled/receives $\mathcal O(\tfrac{1}{\tilde v})$ corrections, we suggest that it should instead be better thought of as a multiple F-string action.

\subsection*{Acknowledgements}

We would like to thank Dario Martelli for discussions and comments. CP is supported by the STFC grant ST/G000395/1.

\bibliographystyle{utphys}
\bibliography{abfuz3}

\end{document}